\documentclass[12pt,epsf]{article}
\usepackage{amssymb,amsmath,amsbsy}
\usepackage{graphicx}
\usepackage{setspace}
\usepackage{subfigure}
\usepackage{cite} 
\newcommand{\be}{\begin{equation}}
\newcommand{\ee}{\end{equation}}
\newcommand{\bea}{\begin{eqnarray}}
\newcommand{\eea}{\end{eqnarray}}
\newcommand{\bear}{\begin{eqnarray}}
\newcommand{\eear}{\end{eqnarray}}
\newcommand{\beas}{\begin{eqnarray*}}

\newcommand{\eeas}{\end{eqnarray*}}
\newcommand{\ba}{\begin{array}}
\newcommand{\ea}{\end{array}}

\newcommand{\ra}

\newcommand{\pd}[2][1]{\ifnum#1=1 \frac{\partial}{\partial {#2}} \else
  \frac{\partial^#1}{\partial {#2}^{#1}}\fi}
\newcommand{\dpd}[2][1]{\ifnum#1=1 \dfrac{\partial}{\partial {#2}} \else
  \frac{\partial^#1}{\partial {#2}^{#1}}\fi}
\newcommand{\td}[2][1]{\ifnum#1=1 \frac{d}{d{#2}} \else
  \frac{d^#1}{d{#2}^{#1}}\fi}

\newcommand{\nbox}{{\,\lower0.9pt\vbox{\hrule \hbox{\vrule height 0.2 cm \hskip 0.19 cm \vrule height 0.2 cm}\hrule}\,}}

\def\href#1#2{#2}

\usepackage{color}

\usepackage{xcolor}
\usepackage[citebordercolor=green, linkbordercolor={ 0 0 1}, linktocpage=true]{hyperref}

\begin{document}
\begin{titlepage}
\begin{NoHyper}
\hfill
\vbox{
    \halign{#\hfil         \cr
           } 
      }  
\vspace*{20mm}
\begin{center}
{\Large \bf New Insights into  Quantum Gravity from Gauge/gravity Duality}

\vspace*{15mm}
\vspace*{1mm}
Netta Engelhardt$^{*}$ and Gary T. Horowitz\\

\vspace*{1cm}
\let\thefootnote\relax\footnote{$^{*}$ Corresponding author: engeln@physics.ucsb.edu, gary@physics.ucsb.edu}

{Department of Physics, University of California\\
Santa Barbara, CA 93106, USA}

\vspace*{1cm}
\end{center}
\begin{abstract}
Using gauge/gravity duality, we deduce several nontrivial consequences of quantum gravity from simple properties of the dual field theory. These include: (1) a version of cosmic censorship, (2) restrictions on evolution through black hole singularities, and (3) the exclusion of certain cosmological bounces. In the classical limit, the latter implies a new singularity theorem.
\vskip 3in
\noindent Essay written for the Gravity Research Foundation 2016 Awards for Essays on Gravitation
\end{abstract}
\end{NoHyper}

\end{titlepage}
\vskip 1cm
\begin{spacing}{1.2}

Curvature singularities are ubiquitous in general relativity. These singularities signal the breakdown of the classical theory: rules governing the evolution of observers subject to arbitrarily large tidal forces should be given by a quantum theory of gravity. Even the near-singularity region defies classical laws, and questions regarding the deep interior of a black hole or very early stages of the universe can only be addressed by a nonperturbative theory of quantum gravity.

While an explicit theory of quantum gravity remains elusive, the celebrated gauge/ gravity duality provides an indirect formulation thereof. Quantum gravity -- specifically string theory -- with asymptotically anti-de Sitter (AdS) boundary conditions is equivalent to a nongravitational quantum field theory (QFT) defined on the conformal boundary~\cite{Mal97}. The string theory is said to live in the `bulk', while the QFT is said to live on the `boundary'. The duality is holographic: it equates quantum gravity with a QFT in a lower dimension.

A powerful aspect of gauge/gravity duality is that statements that are difficult to establish in one theory are often much easier to derive in the dual counterpart. This feature was used to argue for unitarity in black hole evaporation from unitarity of the dual QFT. In this Essay, we employ a similar approach to investigate singularities in quantum gravity. We deduce the following three conclusions about holographic quantum gravity: first, it obeys a version of cosmic censorship; second, it does not allow evolution through black holes to other asymptotic regions; and third, it forbids a large class of cosmological bounces. The final conclusion also leads to a novel singularity theorem in classical gravity. In deriving these conclusions, we make the standard assumptions about gauge/gravity duality (e.g. when the bulk theory has multiple asymptotically AdS regions, each region is associated to a separate QFT; these dual QFTs may be entangled, but they cannot be directly coupled). \\

\begin{figure}[ht]
\centering
\includegraphics[width=8cm]{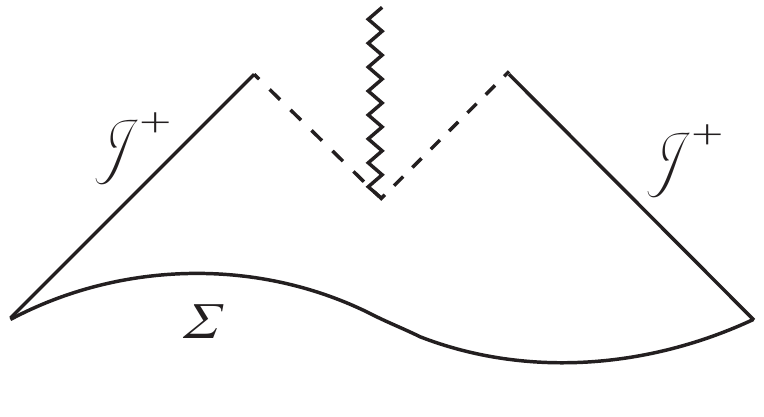}
\caption{Initial data on a Cauchy surface $\Sigma$ evolves to a naked singularity which  prevents evolution to a complete null infinity.}
\label{fig:CosmicCensorship}
\end{figure}

\noindent {\bf Naked singularities:} A longstanding conjecture in general relativity states that generic asymptotically flat initial data has maximal evolution to a complete null infinity~\cite{Pen69, GerHor79}. A proof of this conjecture would rule out naked singularities: these singularities stop unique evolution, as shown in Fig.~\ref{fig:CosmicCensorship}. If this form of cosmic censorship holds in quantum gravity, then naked singularities must be resolved, or alternatively there must be a mechanism in the theory that provides sufficient data at the singularity to fully determine evolution to the future.

In the case of asymptotically AdS initial data, we may use gauge/gravity duality to probe the validity of an AdS analogue of this version of cosmic censorship. Since we do not expect the nature of singularities to strongly depend on asymptotic structure, this may also shed light on naked singularities in spacetimes with different asymptotics. 

Does generic asymptotically AdS initial data evolve to a complete timelike future infinity? The answer provided by gauge/gravity duality is unequivocally yes: with any initial data, the nongravitational QFT has well-defined evolution for all time. The equivalence of the bulk gravitational theory to the dual QFT immediately implies that bulk evolution must continue as well. This in turn indicates that, should naked singularities form in some localized region, deterministic bulk evolution must nonetheless continue for all time. 

Surprisingly, this argument applies even in the classical limit when evolution of the large $N$ gauge theory is well-defined. Classical evolution in this case \textit{must} continue, even in examples of fine-tuned initial data in which naked singularities are known to form (e.g. Choptuik critical collapse of a spherically symmetric scalar field in AdS \cite{Biz11}). The exact mechanism by which evolution continues is unclear, but it is likely that stringy effects (which normally decouple in this limit) remain important near the singularity. The presence of such stringy physics affecting naked singularities was observed in the pinching off of a black string  in~\cite{AhaMarMin04}.

Having established that evolution is not stopped by naked singularities, we turn our attention to other kinds of singularities: those behind a horizon, or at the beginning or end of time -- cosmological singularities. In applying the machinery of gauge/gravity duality to such cases, we invoke a simple fact about field theory: two QFTs that live on two separate spacetimes cannot send signals to one another. We will call this the No Transmission Principle (NTP). When the QFTs in question are holographic, this simple statement has nontrivial consequences for the gravitational dual: the bulk duals to two QFTs on separate spacetimes must be causally disconnected. We discuss the consequences for black holes and cosmologies separately. For more details, see~\cite{EngHor15}.\\

\begin{figure}[t]
\centering
\subfigure[]{
\includegraphics[height=6cm]{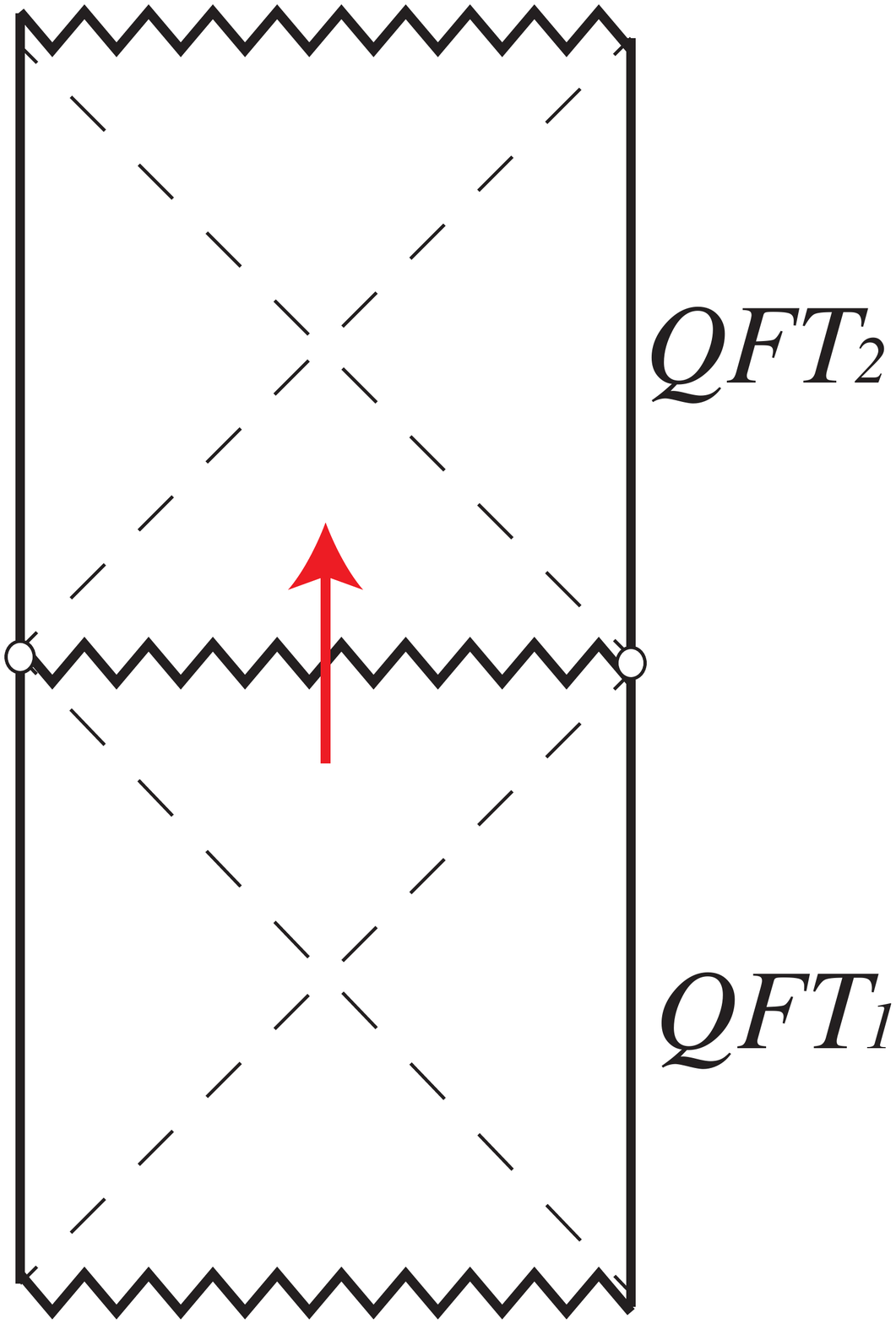}
\label{fig:SchwarzschildEvolution}
}
\hspace{1cm}
\subfigure[]{
\includegraphics[height=6cm]{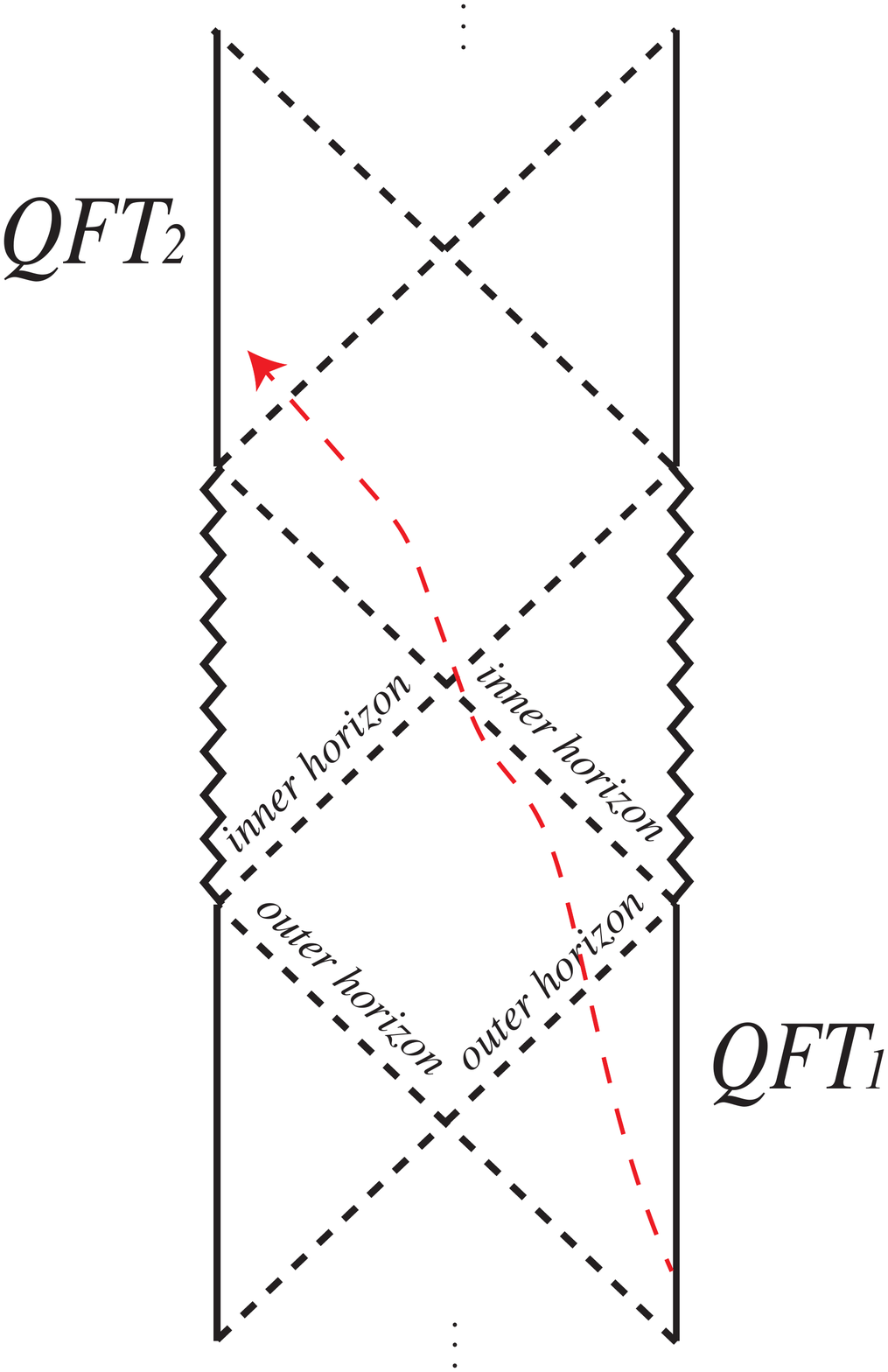}
\label{fig:RNAdS}
}
\caption{\subref{fig:SchwarzschildEvolution} Resolution of the Schwarzschild-AdS black hole singularity allowing evolution to another asymptotic region would violate the No Transmission Principle.  \subref{fig:RNAdS} The possible path of a signal between separate QFTs dual to a charged black hole. Any such signal collapses the inner horizon into a singularity. The No Transmission Principle implies that there is no evolution past this singularity.}
\label{fig:BlackHoles}
\end{figure}

\noindent \textbf{Black holes:} Suppose that quantum gravity could resolve the black hole singularity by allowing evolution to another asymptotic region, as illustrated in Fig.~\ref{fig:SchwarzschildEvolution}. In this case, the black hole singularity would evolve into a white hole of another asymptotic region. The QFT dual to the past of the singularity, QFT$_{1}$, could then send signals to the QFT dual to the future, QFT$_{2}$, through the bulk. QFT$_{1}$ and QFT$_{2}$, however, live on separate copies of the Einstein Static Universe; the NTP forbids signal exchange between them. This immediately implies that evolution through the black hole singularity to another asymptotic region is forbidden.

%
%

The Reissner-Nordstrom-AdS black hole may appear to contradict our conclusions. The  classical solution has multiple asymptotic regions which are connected through the interior of the black hole, see Fig.~\ref{fig:RNAdS}. The inner horizon, however, is known to be unstable: perturbations of the black hole will cause the inner horizon to become singular (as originally noticed in \cite{SimPen73}). It follows from the NTP that no signal can pass through this singularity -- even in quantum gravity.\\

\noindent \textbf{Cosmology:} A cosmological bounce is a hypothesis concerning the quantum gravity resolution of cosmological singularities. In one such scenario, a big crunch singularity -- the end of spacetime -- evolves into a big bang singularity of another universe. In quantum gravity with asymptotically AdS boundary conditions, we now argue that the NTP forbids cosmological bounces.

The crux of the argument is a reversal of the logic used in favor of cosmic censorship: in the latter case, we argued that evolution in the bulk must continue when evolution of the dual field theory does. Some QFTs are singular in the sense that they cannot be evolved past a certain time; for example, when the QFT lives on a spacetime which itself has a cosmological singularity.\footnote{When the QFT is conformally invariant, one can define arrested evolution by working in a \textit{standard conformal frame} -- a frame in which the QFT spacetime is maximally conformally extended and spatially compact, and where the volume of certain spatial slices is bounded from above and below. The notion that QFT evolution ends is now conformally invariant.}  When QFT evolution stops, so must evolution in the dual bulk: the bulk must develop a cosmological singularity. 

\begin{figure}[t]
\centering
\subfigure[]{
\includegraphics[height=6cm]{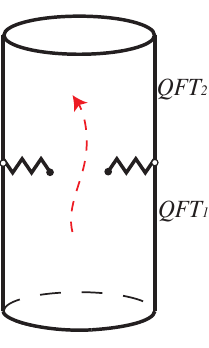}
\label{subfig:WithHole}
}
\hspace{1cm}
\subfigure[]{
\includegraphics[height=6cm]{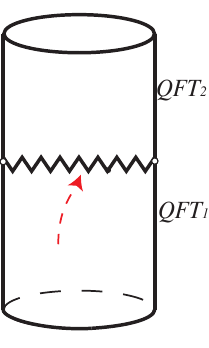}
\label{subfig:WithoutHole}
}
\caption{\subref{subfig:WithHole} If the bulk singularity allows evolution, signals can travel from the bulk region dual to QFT$_{1}$ to the region dual to QFT$_{2}$, which is forbidden by the No Transmission Principle.  \subref{subfig:WithoutHole} The NTP implies that any signal in the bulk region dual to QFT$_{1}$ must terminate before it reaches the region dual to QFT$_{2}$.}
\label{fig:SingularityWithHole}
\end{figure}

In a model in which this cosmological singularity is resolved via a bounce to another region, the resulting bulk has two dual QFTs: a past-singular QFT and a future-singular QFT. A bouncing bulk would thus facilitate signal transmission between these two separate QFTs, in violation of the NTP (see Fig.~\ref{subfig:WithHole}). Bulk evolution must therefore stop at finite time, as in Fig.~\ref{subfig:WithoutHole}; holographic bounces are  forbidden\footnote{We are only able to exclude bounces when the dual field theory is singular. We are aware of only one example of a bulk cosmological singularity for which the dual field theory may be nonsingular \cite{AwaDas08}.}.

At the classical level, the result above implies a new type of singularity theorem: if a supergravity solution arising in the low-energy limit of string theory has (1) asymptotically AdS boundary conditions, and (2) a cosmological singularity on the boundary, then the bulk solution must also have a cosmological singularity.

We close with a cautionary remark. Our conclusions could be avoided by associating a state in one QFT with a state in another, but there is no natural way to do so. Such  matching requires the existence of a limiting state in one of the QFTs, and this limit often does not exist. In the case of the black hole, the dual QFT evolves indefinitely, and the microstate never settles down. The cosmology dual is similar: a singular QFT is unlikely to have a well-defined limiting state.  One could make up an arbitrary rule to identify the states, but this rule would contain extra physics necessary to describe the bulk that is not  captured by the dual field theory. This contradicts the statement that the two dual descriptions are equivalent.

In summary, we have shown that gauge/gravity duality provides significant insight into the treatment of singularities in quantum gravity. Naked singularities are either resolved or endowed with data that allows evolution through them, while black hole and cosmological singularities remain a true end to spacetime. Further study of gauge/gravity duality is likely to yield more insights into the mysteries of quantum gravity.

\end{spacing}
\section*{Acknowlegements}
It is a pleasure to thank M. Dafermos, D. Marolf, S. Minwalla, and H. Reall for discussions. 
 This work was supported in part by NSF grant PHY-1504541. The work of NE was also supported by the NSF Graduate Research Fellowship under grant DE-1144085 and by funds from the University of California. 

\bibliographystyle{JHEP}

\bibliography{all2}

\end{document}